\documentclass[sigplan,10pt,nonacm]{acmart}
\usepackage[export]{adjustbox}
\usepackage{comment}
\usepackage[utf8]{inputenc}
\usepackage[T1]{fontenc}
\usepackage{epsfig} 
\usepackage{multirow}
\usepackage{amsmath} 
\usepackage{listings}
\usepackage{subfig}
\usepackage{svg}
\usepackage{xcolor}
\usepackage{fancyhdr}
\usepackage{enumitem}
\setlist[itemize]{align=parleft,left=0pt..1em}
\usepackage{tikz}
\usepackage[frozencache,cachedir=.]{minted}
\usepackage{xspace}

\AtBeginDocument{%
  \providecommand\BibTeX{{%
    \normalfont B\kern-0.5em{\scshape i\kern-0.25em b}\kern-0.8em\TeX}}}

\setcopyright{acmcopyright}
\copyrightyear{2018}
\acmYear{2018}
\acmDOI{10.1145/1122445.1122456}

\acmConference[Woodstock '18]{Woodstock '18: ACM Symposium on Neural
  Gaze Detection}{June 03--05, 2018}{Woodstock, NY}
\acmBooktitle{Woodstock '18: ACM Symposium on Neural Gaze Detection,
  June 03--05, 2018, Woodstock, NY}
\acmPrice{15.00}
\acmISBN{978-1-4503-XXXX-X/18/06}

\newcommand{\pnlname}{{\em PsyNeuLink}}
\newcommand{\sysname}{{\em Distill}\xspace}

\title{{\em \sysname}: Domain-Specific Compilation for Cognitive Models}

\begin{document}

\author{J\'{a}n Vesel\'{y}}
\authornote{Joint first authors.}
\affiliation{%
  \institution{Yale University}
    \country{USA}
  }
\author{Raghavendra Pradyumna Pothukuchi}
\authornotemark[1]
\affiliation{%
  \institution{Yale University}
    \country{USA}
}
\author{Ketaki Joshi}
\affiliation{%
  \institution{Yale University}
    \country{USA}
}
\author{Samyak Gupta}
\affiliation{%
  \institution{Princeton University}
  \country{USA}
}
\author{Jonathan D. Cohen}
\affiliation{%
  \institution{Princeton University}
  \country{USA}
}
\author{Abhishek Bhattacharjee}
\affiliation{%
  \institution{Yale University}
    \country{USA}
}
\begin{abstract}
This paper discusses our proposal and implementation of {\em \sysname}, a domain-specific compilation tool based on LLVM to accelerate cognitive models.
Cognitive models explain the process of cognitive function and offer a path to human-like artificial intelligence. However, cognitive modeling is laborious, requiring composition of many types of computational tasks, and suffers from poor performance as it relies on high-level languages like Python. In order to continue enjoying the flexibility of Python 
while achieving high performance, {\em \sysname} uses domain-specific knowledge 
to compile Python-based cognitive models into LLVM IR, carefully stripping away features like dynamic typing and memory management that add overheads to the actual model. As we show, this permits significantly faster model execution. We also show that the code so generated  enables using classical compiler data flow analysis passes to reveal properties about data flow in cognitive models that are useful to cognitive scientists. 
{\em \sysname} is publicly available, is being used by researchers in cognitive science, and has led to patches that are currently being evaluated for integration into mainline LLVM.
\end{abstract}

\maketitle

\section{Introduction}
\label{sec:introduction}

Cognitive scientists use computational models to understand how humans achieve cognition and how the brain gives rise to the mind.
Cognitive models simulate mental processes by describing how inputs (e.g., visual or auditory stimuli to the brain) are processed by a set of interconnected 
subsystems to generate 
behavior.
Insights from cognitive modeling have greatly influenced not only the brain, psychological, and cognitive sciences, but also the field of artificial intelligence (AI).
In fact, this influence has been instrumental in shaping AI right from the onset of artificial neural networks to the recent advances in deep learning for gameplay and scientific computing from DeepMind~\cite{kumaran:cls, jumper:alphafold}.

Researchers use cognitive models to obtain distributions of outcomes and their evolution over time for various cognitive tasks.
By construction, these models are designed to be run hundreds of thousands of times because of inherent stochasticity, in order to build histograms of outcomes across multiple parameter settings, and/or to assess the dynamics of cognitive processes across a series of time steps. 

The vast majority of cognitive models are developed in Python-based frameworks~\cite{psyneulink,pytorch,hines:neuron}. This is because Python eases programming significantly, lowers the barrier to entry to computational methods, is amenable to rapid prototyping, and offers access to optimized scientific computing libraries \cite{scipy, numpy}.
But, the use of Python also introduces performance inefficiencies. As more sophisticated cognitive models are built to capture advanced brain processes, these performance inefficiencies worsen to the point where cognitive models take several days to weeks to run. As a consequence, scientific progress is impeded by inefficient modeling tools.


One might consider alleviating the slow runtime of cognitive models via publicly-available compilation tools like PyPy~\cite{pypy} and Pyston~\cite{pyston}.
In practice, however, we find that these approaches leave performance on the table. The central problem is that Pypy and Pyston cannot optimize complex dependencies in cognitive models because of the runtime checks needed to deal with Python's dynamic data structures and dynamic typing. These features also obscure the natural parallelism available in cognitive models, and impede the ability to  offload portions of the models onto hardware accelerators for which they are otherwise suitable. Adding to these challenges is the fact that large-scale cognitive models increasingly require integration of separate subsystems developed across a range of different environments (e.g., PyTorch~\cite{pytorch}, Emergent~\cite{aisa:emergent}, NEURON~\cite{hines:neuron} or PsyneuLink~\cite{psyneulink}); it is difficult for compilers to optimize across computations expressed in this multitude of environments.

Yet another possibility might be to build a domain-specific language (DSL) targeted at cognitive scientists. While this is likely to enable maximal performance, it also requires large-scale community buy-in and porting of many models already built across many research institutions using Python. Even if these practicality concerns could be surmounted, the extreme heterogeneity of the cognitive modeling ecosystem presents major roadblocks to designing a canonical set of language constructs and software tools needed for a domain-specific language. This scale of heterogeneity can be appreciated by realizing that cognitive models can integrate components
with varying levels of biological fidelity, developed by different frameworks and research groups; e.g., a single model can include neurally accurate descriptions of some brain structures, an artificial neural network from machine learning to determine the attention allocated to inputs and a behavioral model of control to modulate the pathways.

In response, we build \sysname{}, a dynamic compilation tool for cognitive models that exploits domain knowledge to generate efficient code. \sysname{} uses domain specific knowledge from cognitive science to aggressively eliminate Python's dynamic code, and generates LLVM IR for all the components in a model, including those developed in ancillary frameworks (e.g., Pytorch). This approach is predicated on the observation that, like practitioners in other scientific computing communities~\cite{weld, tuplex}, cognitive scientists use Python for its flexibility and access to optimized scientific computing libraries~\cite{scipy, numpy}, but do not need many of the dynamic language features that slow execution down. Once these dynamic features are removed, standard optimizations on LLVM IR yield notable speedups, and permit leveraging LLVM's broad existing family of code generation backends (with no change) for multiple CPU architectures and accelerators. We find that \sysname{} accelerates cognitive model execution by an average of 26$\times$ and maximum of 778$\times$ across our evaluated cognitive models compared to Pyston and PyPy.
A model that couldn't complete even in 24 hours could finish in less than 5 seconds when \sysname{} was used. \sysname{} can also extract parallelism from the models and target multi-core CPUs and GPUs, resulting in additional 4.8$\times$ and 6.4$\times$ speedups, respectively. 

In the process of accelerating model execution, we were also intrigued by another (unexpected) benefit of stripping away Python's dynamism from LLVM IR. All cognitive models are expressed as computational graphs, and we found that eliding Python's dynamism enabled \sysname{} to produce control and dataflow graphs (CDFGs) of LLVM IR with similar shape and data flow as the original model. We leverage this observation to augment \sysname{} with compiler analyses for deducing high-level semantic information from the CDFG. This in turn permits \sysname{} to automate several types of model-level analysis that have traditionally been undertaken manually and tediously by scientists.
It also enables \sysname{} to discover user-guided optimizations specific to cognitive models. 

We demonstrate two important applications of these user-guided analyses and optimizations. First, \sysname{} identifies cases where entire models can be verified to be equivalent, and also recognizes that certain complex nodes are equivalent with, and hence, can be replaced by simpler modules that have an analytical solution.
Second, \sysname{} calculates the impact of a cognitive model's parameters on the outputs and finds their optimal values entirely with compiler analysis techniques that extend LLVM's range value propagation and scalar evolution passes. As this parameter estimation process was performed manually over hundreds to thousands of runs, \sysname{}'s automation of this step saves days to weeks of modeling effort. Moreover, because of the general utility of our enhancements to LLVM's passes (i.e., extending support for integers to floating point), we have submitted a patch to the LLVM community for mainline integration.

\sysname{} is undergirded by three main design principles. First, we wish to avoid requiring cognitive scientists to change the source-code of their models or frameworks. Second, we delegate performance extraction to the compiler, allowing scientists to focus on creating models in the manner most intuitive to them. Third, we also minimize software engineering effort, development, and maintenance cost to compile the models. This last goal is central to our decision to reuse LLVM IR and its associated infrastructure to build \sysname{}, and avoid the need for a new DSL or IR. More specifically, our contributions are:

\begin{enumerate}
    \item \sysname{}, a compilation tool that exploits domain-specific knowledge to provide near-native execution speeds for cognitive models along with support to offload computations on accelerators. \sysname{} does not require changes to source code and reuses existing LLVM infrastructure.
    \item Identifying that user-guided analyses and optimization can be performed by compiler analysis, and incorporating them into \sysname{}.
    \item An evaluation of \sysname{}-accelerated models on single and multicore CPUs and GPU.
\end{enumerate}

\sysname{} is publicly available and is being used in several leading research labs via its plug-in with PsyNeuLink~\cite{psyneulink}, a newly-developed framework that can import and run sub-models developed in various environments.
\sysname{} enables the design of larger and more complex cognitive models than previously possible. This is an important and necessary step towards the larger goal of understanding human cognition and replicating its processes in artificial intelligence.   

\section{Background and Motivation}
\label{sec:background}

\subsection{Cognitive Models Structure and Computation}
\label{sub_cogmodel}

Cognitive models are expressed as computational graphs and represent how neural or psychological components process  inputs (typically sensory stimuli, but also potentially non-sensory stimuli like memory) to produce behavior.
These models are used for several purposes like fitting experimental data, simulating a cognitive process, producing an idealized outcome or for what-if analysis to understand the impact of tunable structures and parameters.

Cognitive models can be represented as a graph where the nodes are sub-processes or computational functions involved in the overall task, and the edges represent projections of signals between nodes. Nodes perform their computation when their activation conditions are met (e.g., the appearance of an input, the passing of a specified time period). 

\begin{figure}[ht]
    \centering
    \includegraphics[width=0.8\linewidth]{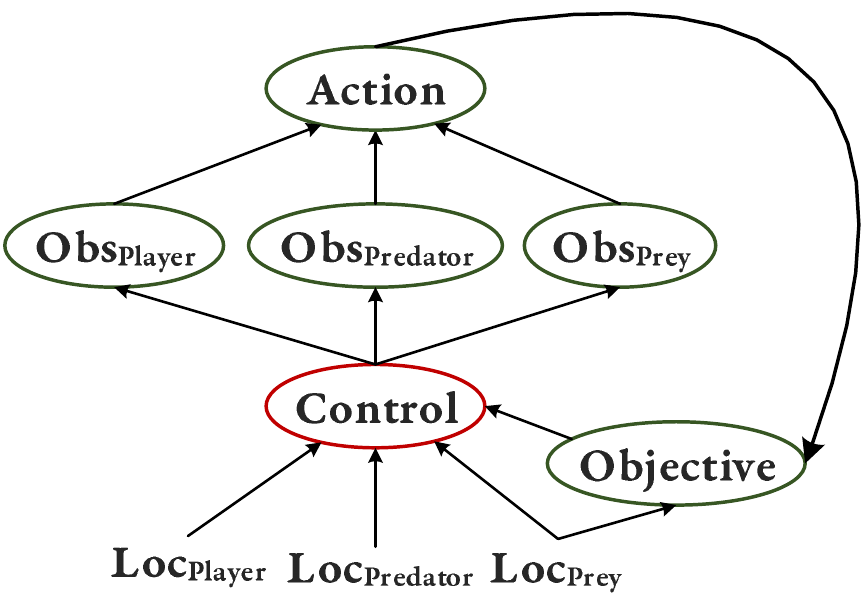}

     \caption{A cognitive model of an agent performing the predator-prey task.}
     \label{fig:pp_eg}
\end{figure}

We illustrate a simple cognitive model using the well-known predator-prey task \cite{willke2019comparison}.
In this task, an intelligent agent, either a human or non-human primate, is given a joystick and shown a screen with three entities -- a player whose position is controlled using the joystick, a prey that the player must capture, and a predator that the player must avoid.
Figure~\ref{fig:pp_eg} shows a cognitive model to study the role of attention in the agent's performance on this task.
The agent's attention is limited and there is a cost of paying attention to an entity.
Attention paid to an entity determines the accuracy with which the agent can see that entity's location. Moreover, the agent does not have to distribute its attention fully.

The \textit{Action} node calculates the player's movement based on the observed positions of the three on-screen entities.
The role of attention is modeled through the \textit{Control}, \textit{Objective} and \textit{Obs} nodes. \textit{Control} takes the exact 2-dimensional (2D) coordinates of all the entities (\textit{Loc} values) and allocates attention to each of them.  This allocation determines the variances of three 2-D gaussian distributions whose means are the actual locations of the respective entities.
These distributions are then sent to the \textit{Obs} nodes that sample from them to generate the observed locations.

To determine the best attention allocation, the \textit{Control} node searches over all possible attention allocations, evaluating the cost of each allocation and the quality of the associated move. 
The cost of each allocation is calculated by \textit{Control}, and the quality of the move is computed by the \textit{Objective} node.
The \textit{Objective} node uses the direction given by \textit{Action} for an allocation and the true location of the prey to compute the goodness of the move. The \textit{Control} node then selects the parameters that have the lowest cost. Overall, the entire process from reading inputs and searching over allocations is repeated for each time step until the prey or the player is captured. 
Cognitive scientists are interested in both the final outcomes and the dynamics over time. 


Figure \ref{fig:pp_eg} shows the basic predator-prey model form, but advanced variants of the model can also include cognitive processes for memory in order to recall previous locations, neural networks trained on experimental data to produce a move, or, visual processors to extract locations from screen frames. Components pertaining to these augmentations can come from different frameworks like PyTorch, Emergent or NEURON~\cite{hines:neuron,aisa:emergent,pytorch}.
Cognitive models may also include state that is modified across each set of inputs (e.g., the cost of the previous allocation), capping the available parallelism for execution. Model nodes also usually store metadata to track information like the number of times a particular node has executed, as this information is useful and of interest to scientists. 


\subsection{Cognitive Model Execution} 
\label{sub_run}

While cognitive models have historically been developed in a variety of environments, recent efforts in the community focus on converging on a single "lingua franca" environment that can support models built in others \cite{mdf-grant}. PsyNeuLink is prominent among emerging standardized environments, and is hence the focus of our work. Listing~\ref{listing:sched} shows how cognitive models execute in PsyNeuLink. 

First, in a sanitization check to ensure that nodes are properly connected, the framework runs through all nodes, initializing all parameters and inputs with default values and propagating inter-node signals. The shapes of each node's inputs and outputs in the sanitization run must match those that will be used in the actual run.  

Second, scheduler logic within PsyNeuLink reads the set of all inputs with which the model has to be run. The variable \texttt{inputs} in Listing~\ref{listing:sched} is a list, where each element consists of a set of values that are input to the model (e.g., the locations at one time step for the predator-prey model). A single model may have multiple inputs.

Third, the scheduler runs the model for a number of trials (TOTAL\_TRIALS in the listing), where each trial runs the model with one input. A trial consists of multiple iterations of model execution. In each iteration, the scheduler identifies nodes that are ready to run, and then executes them. The scheduler identifies the nodes that are ready to run by relying on the activation conditions that are explicitly specified per node. Examples of such conditions include waiting until other nodes are run a certain number of times, until the outputs  of particular nodes stabilize, or after a certain amount of time has elapsed. 

\begin{listing}[ht]
\centering
\begin{minted}[frame=lines,framesep=2mm,
baselinestretch=1.2,
fontsize=\footnotesize,
linenos]{python}

#Sanity check the model
sanitize(model)

#Read all inputs
inputs = read_inputs()

#Begin a model's run
num_trial = 0
while num_trial < TOTAL_TRIALS:
    #Each trial reads an input
    input = inputs[num_trial % len(inputs)]
    #Run scheduling iterations
    while not end_of_trial:
        find_ready_nodes(model)
        run_ready_nodes(model, input)
    num_trial += 1
\end{minted}
\caption{Outline of the scheduling loops in a framework like PsyNeuLink that runs cognitive models.}
\label{listing:sched}
\end{listing}

In sum, apart from running nodes from different frameworks, execution switches between nodes and the scheduling logic. These back and forth transitions between computation and scheduling impacts performance, motivating the need to accelerate complex and composited models.

\subsection{Shortcomings of Dynamic Compilation  Tools}

Dynamic compilation is a popular approach to accelerate applications written in high-level languages like Python. Indeed, Pyston results in average execution time improvements of 43\% for a group of cognitive models we test. However, they still leave significant opportunities for optimization. First, they cannot easily identify and leverage opportunities to reduce runtime overheads by tracking model control flow. For example, the predator-prey model in Section~\ref{sub_cogmodel} is run many times for a single input, but the path of execution is the same for all these runs. This is typical of cognitive models, but takes significant time and space resources for PyPy and Pyston to track.

Second, existing dynamic compilation tools do not fully eliminate dynamic features of Python unnecessary for the model. As an example, the signals between the nodes of a cognitive model not only have a fixed type, but also have a fixed shape across runs. Therefore, dynamic Python structures such as lists and dictionaries that are used to hold these values can be safely compiled to static data structures. However, existing tools conservatively conform to Python semantics, retaining these unnecessary computations.  

Third, Pyston and PyPy cannot optimize across computations from different frameworks, and across the scheduling invocations between executions of the model nodes. When a model uses computations from multiple environments like PyTorch and PsyNeuLink, even if the separate components are compiled, optimization does not cross these frameworks. Additionally, execution frequently switches between the computations in the nodes and the scheduling logic in the modeling framework that identifies which nodes are ready to run. This also limits the scope of optimizations and results in the execution switching between compiled and interpreted modes, compromising performance. 

Finally, available dynamic compilation tools cannot automatically extract parallelism from the models or offload computations to accelerators like GPUs. This is a wasted opportunity because cognitive models are run many times, and there are several dimensions along which computations can be parallelized. For example, in the predator-prey model, the evaluations for each combination of attention allocations could have been run in parallel. Furthermore, when multiple samples are drawn from the distributions of observed location, each sample and subsequent action could also be computed in parallel. While one might consider leveraging existing multithreading and GPU programming libraries for Python, they all require scientists to explicitly identify such parallel computations and mark functions to be offloaded to a GPU. A more desirable solution is to automate these steps so that cognitive scientists can solely focus on their designs rather than grapple with parallel programming constructs.

The confluence of these shortcomings leads to cascading slowdowns in the execution. For example, unoptimized data structures not only have longer access times, but also prevent subsequent optimization passes by hindering the propagation of values and references.  Moreover, multithreading with Python does not result in parallel execution because the threads are serialized by the Global Interpreter Lock~\cite{beazley2010understanding}, unless the threads run compiled code, in which case they do not have to take the lock. Thus, to maximally benefit from parallelism, it is important to compile the Python threads.





\section{\sysname{}: Domain-Specific Compilation for Cognitive Models}
\label{design}

Cognitive models are graphs of computations expressed in various environments, with complex scheduling rules for execution. The models are constructed in Python but can be composed from heterogeneous sub-models developed in multiple frameworks.
Unfortunately, dynamic compilation with existing tools is not effective.
\sysname{} uses domain-specific knowledge of cognitive neuroscience to aggressively optimize the models to yield near-native execution speeds. 

We observe that while Python's dynamic features make it easier to construct the models, they are not necessary to run them.
Therefore, \sysname{} aggressively eliminates dynamic features in the models and converts dynamic data structures to statically defined ones, yielding substantial acceleration.
Furthermore, \sysname{} automatically extracts parallelism and computations that are offloaded to GPUs.

\sysname{} plugs into PsyNeuLink~\cite{psyneulink}, which is a newly-developed framework that can import and run sub-models developed in various environments.
In this section we describe the steps in \sysname{}'s code transformations and parallelism extraction to accelerate cognitive models. 

\subsection{Type and Shape Extraction}
\label{sub_types}

 As described in Section~\ref{sub_cogmodel}, repeated computations in the models have the same types and shapes.
 Thus, the first step in compiling cognitive models is to deduce the types and shapes of the values used in the computations.
 Fortunately, this information is readily available from the sanitization pass run by the framework (Section~\ref{sub_run}).
 By construction, the value types and shapes of the inputs and outputs of each node in this run must be the same as what will be used in the actual runs.
 \sysname{} uses this information to infer the types and shapes of all computations in the model. 
 
 \subsection{Identifying Code Paths for Compilation}
 \label{sub_hotpath}
 
 Typically, dynamic compilation requires expensive analysis and tracking to determine which code paths need to compiled and when. In our case, the main computations of each node are in an \texttt{execute} method in the nodes, and we know that these computations must be compiled because they are repeatedly invoked and are time-consuming. Thus, we do not need to run expensive dynamic hot path analysis.
 We leave code intended for initialization and visualization because they are not repeatedly invoked.
 
 \subsection{Dynamic to Static Data Structure Conversion}
 \label{sub_dynstruct}
 
Cognitive models use dynamic Python data structures like dictionaries and lists for node inputs, parameters, and outputs. We observe that their shapes (and keys in the case of dictionaries) remain invariant during execution. We therefore convert these entities into statically-defined structures. More specifically, we make the following changes:
 
First, we create two structures that hold the values of the outputs of all nodes in the current and previous iterations (see the inner loop of Listing~\ref{listing:sched}). Node outputs are written to these top-level structures. We need two structures because multiple nodes running in the same iteration consume the values created in the previous iteration.  
 
Next, we create separate structures for read-only and read-write parameters in the models. Parameters exist at the node level (like the attention levels in the \textit{Control} node of the predator-prey model), or as arguments of functions (e.g., the amplitude argument of a function that computes a sinusoid). Such functions are defined in the nodes or the standard framework library.

Creating separate structures will be useful when we parallelize executions, where the threads only need to make local copies of the read-write parameter structure.

Modeling frameworks additionally contain two structures, one for the set of inputs for all trials (the variable \texttt{inputs} in Listing~\ref{listing:sched}) and another for overall outputs in the trials. We convert these two entities into arrays.
 
Finally, the original computations commonly use strings as keys to fetch data, and we convert these strings to enumerated entries (enums), that are used as offsets to index into the structures.

The information about the sizes and keys of the model's parameters and outputs is available in the sanitization run, and for the inputs, this information is available whenever they are read. In scenarios where \sysname cannot infer the shapes of some intermediate variables statically, it does not compile the models. However, we haven't seen such a case in the cognitive models we tested.

In the scenario where a model selects the parameter configuration with the minimal cost (e.g., attention allocation in the predator-prey model), it is possible that multiple parameters may give the same minimal cost. In such cases, it is customary to randomly pick one parameter choice. To implement such constructs, we use reservoir sampling~\cite{vitter1985random} so that we do not need to store a variable number of potential parameter choices, and then choose one from that list. With reservoir sampling, we can have a fixed size statically defined datastructure.

 Eliminating dynamic datastructures  significantly reduces their access times, and enables several optimizations. In their new format, it is easy to propagate values and references for subsequent optimizations. In addition, the datastructures are now compact, improving cache performance. 

\subsection{Generating LLVM IR}
\label{sub_llvm}

\sysname generates LLVM IR for all the nodes and their functions before the model begins execution. 

\subsubsection{Code Specialization}
\label{subsub_spl}

For the functions in the framework's standard library, there are pre-defined templates which are then specialized to the types with which they are called.
In the original model, a single function could have been invoked with different types of parameters due to Python's polymoprhic semantics.
\sysname{} generates monomorphic code, creating a separate version of the function for each lexical instance it is invoked.

All the nodes also have a generic template with the basic structure of a node that is filled with the computations in the model's nodes.

\subsubsection{Generating Code for Multiple Libraries and Frameworks}
\label{subsub_codegen}

Recall that cognitive models can use computations from other libraries and frameworks.
\sysname{} takes these computations commonly used in cognitive models and generates LLVM IR.
This includes simple functions from the numpy library like the logistic and sigmoid functions, and neural networks and optimizers from PyTorch.
Lowering these different computations to a common IR allows optimization to span across them, resulting in more efficient code.

\subsection{Optimizations}
\label{sub_optim}

 After \sysname generates the LLVM IR for a model, we run LLVM's standard optimization passes. These optimizations like constant propagation and loop invariant code motion can work across computations from multiple frameworks, and across computations from the model and its scheduler to create optimized code. For example, these optimizations could identify that when the \textit{Control} node in the predator-prey model creates an attention allocation, the \textit{Obs} nodes can be run without an explicit check by the scheduler to see which nodes are ready.
 
 Additionally, generating a common IR by itself removes the invocation of the Python interpreter during the entire course of model execution, significantly improving performance.

Lastly, the compiled code in Python does not hold the Global Interpreter Lock, enabling true parallel execution when multithreading is used.


\subsection{Parallelism and GPU Acceleration}
\label{sub_parallel}

When a cognitive model contains a node that evaluates parameters using exhaustive grid search (e.g., the \textit{Control} node in the predator-prey model), each evaluation can be run in parallel. 
In this case, \sysname is capable of automatically extracting multicore parallelism and offloading computations to accelerators like GPUs. 

To generate multi-threaded code, \sysname first creates as many Python threads as the available cores. Each thread is assigned a segment of the grid search space, and evaluates the parameters in this space using functions that have been compiled in the previous step. Each thread also maintains a local copy of the read-write parameters and the node outputs that it writes to. Since the threads only run compiled code, they do not take the Global Interpreter Lock and can be run in parallel.

For GPUs, we use the NVPTX backend included with LLVM~\cite{holewinski2011ptx} to generate the NVPTX IR for the evaluations. This process generates a kernel for the evaluation function where each thread evaluates one point in the grid search space. The generated kernel is imported to CUDA by PyCuda~\cite{klockner2010pycuda} and executed on GPUs.

For reproducibility, models that sample from random number generators (e.g, sampling the location distributions in the predator-prey model) use independent random number generators for all evaluations.
In these models, the state of the pseudo-random number generator (PRNG) is used as a read-write parameter in their evaluation functions, which is used and restored on every invocation.
This approach directly allows even multiple threads running in parallel to draw the same random numbers. However, as we will see later, replicating the state for each invocation has significant storage overheads.

\subsection{Putting It All Together}
\label{sub_summary}

\sysname aggressively eliminates dynamic features used in cognitive models, and generates LLVM IR for the computations in cognitive models, even if they come from different environments. This allows model-wide optimizations, avoids invoking the interpreter and enables true parallelism. Existing tools like PyPy and Pyston do not perform these optimizations, leaving models to run inefficiently.
\section{Augmenting \sysname with Model Analysis}

We observed that the CDFG of the LLVM IR generated by \sysname for a model matches closely with the interconnection of nodes in the model. This inspired us to augment \sysname with compiler analysis that can provide model-level information to users. 
Analyzing a model's outcome when parameters or nodes are modified in a model is an important aspect of cognitive modeling. For example, a researcher may want to know what happens to the overall objective in the predator-prey model if a `fear amplifier' node that modulates the attention allocated to the predator is added to the model.

We have discovered that we could perform several such analyses entirely in the compiler by suitably modifying the analyses present with the LLVM infrastructure. This means that it is not necessary to run the models at all to obtain such information, saving significant resources.

Furthermore, such model-level analyses enables user-guided optimizations where \sysname can present users (cognitive scientists) with multiple code generation alternatives, each of which has slightly different numerical properties. 

Next, we describe these model analyses and optimizations, their significance and the changes we made to LLVM to support them.

\subsection{Sensitivity to Parameter Values}
\label{sub_sens}

A common case in cognitive modeling is identifying the impact on outputs for different choices of parameter values. Typically, this is done by running the model with all the choices of these parameters. However, we can perform such analysis entirely in the compiler using value range propagation.

Value range propagation (VRP) is a dataflow analysis that determines ranges of variables based on control flow, type restrictions, and used operations.
For example it can determine that $exp(x)$ can only ever by a positive number or $NaN$, and a commonly used {\em Logistic} function can be shown to always output values in the range (0,1]. However, LLVM implements VRP only for integers. Therefore, we extend this implementation to  support floating point types and common floating point operations.

Extending VRP to floating point ranges is useful beyond analyzing cognitive models.
Many floating point operations need special handling in the presence of special values like negative zero, not-a-number, or infinities.
While the compiler can be instructed to optimize these using special fast-math optimization flags,
these are currently set globally per compilation unit or per function, or tracked in a limited way.
Floating point ranges can be used to determine the absence of such special values for each operation and fast-math optimizations can be applied without breaking strict semantics.
This is especially useful for GPU targets which often have specialized fast instructions that do not fully adhere to IEEE floating point semantics.

\subsection{Estimating Convergence times}
\label{sub_est}

For models that simulate accumulation of evidence over time, cognitive scientists are interested to know the estimated time by which evidence accumulation leads to a decision, when they vary the model parameters. 
We can perform such analyses using the scalar evolution pass with LLVM. Scalar evolution (SCEV) extends VRP to loops to track value ranges across loop iterations and calculating the number of loop iterations if possible. Similar to VRP, we extend LLVM's SCEV pass to support floating point types and to calculate the minimum number of loop iterations.
Variable ranges at a loop exit can be used to continue range analysis beyond loops.

\subsection{Adaptive Mesh Refinement for Subspace Search}
\label{sub_amr}

In cognitive models that search in a parameter space, it is useful to know which regions of the space results in noteworthy behavior from the model. We can use VRP to progressively narrow the sub-space of interest akin to adaptive mesh refinement. In fact, it is possible to estimate the best parameters without running the model. 

Consider the predator-prey model. This model uses grid search to find the best attention allocation. For simplicity, consider that we want to find the best attention allocation for the prey when a fixed attention allocated to the predator and player. The conventional approach is to run the model for various levels of attention (among 100 possible levels) for the prey. Moreover, for each level, the model must be run several times to obtain the output distribution.

\begin{figure}[h]
\includegraphics[width=\linewidth]{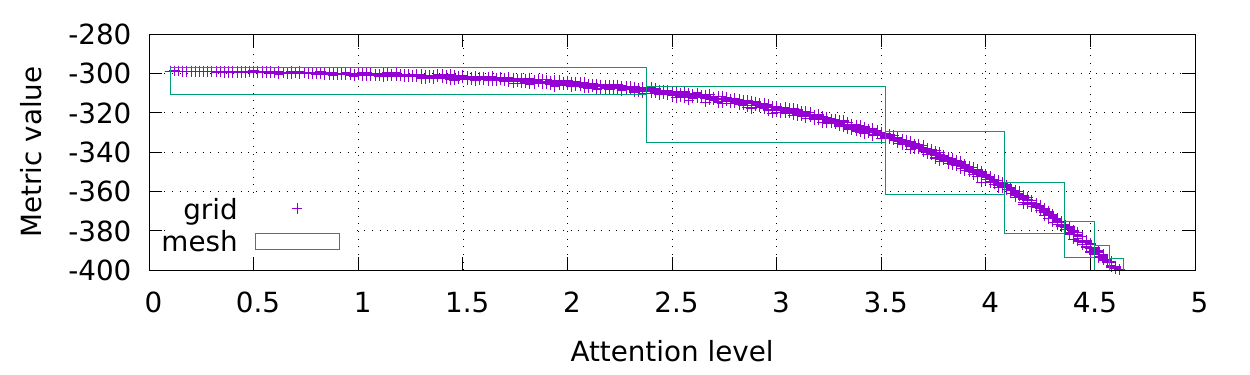}
\caption {Search for minimum using mesh refining in the Predator-Prey model superimposed over sampled grid}
\label{fig:predator-amr}
\vspace{-2mm}
\end{figure}

Figure~\ref{fig:predator-amr} shows how VRP and binary search can be used to find the optimal attention allocation to prey through compiler analysis. The X-axis of the chart is the attention level allocated to the prey and the Y-axis is a cost metric to evaluate the allocation. The boxes show how the search space is progressively refined. For example, the first iteration of the analysis finds that the range of the metric's value is lower in the region between 2.4  and 4.6, than in the region from 0 to 2.4. Therefore, it performs another binary search to find the metric's value in the region from 2.4 to 4.6 and so on. Eventually it finds the optimal allocation, which is close to 4.6 after about 7 rounds. To determine the same outcome by running the model required hundreds of thousands of runs (also shown in the chart). This shows how VRP can be a great tool for cognitive scientists.

\subsection{Clone detection}
\label{sub_clone}

There are multiple benefits of knowing if a node or an entire model is equivalent to another. In the simplest case, it helps verify correctness when a researcher creates an alternate version of the model,  without changing the computations. This can happen when models are changed to be more intuitive without affecting their computational behavior. Another use is where a complex node can be substituted with an equivalent whose computation is simpler.

\begin{figure}[ht]
\centering
\subfloat[Leaky Competing Accumulator (LCA)]{
\includegraphics[width=0.45\linewidth]{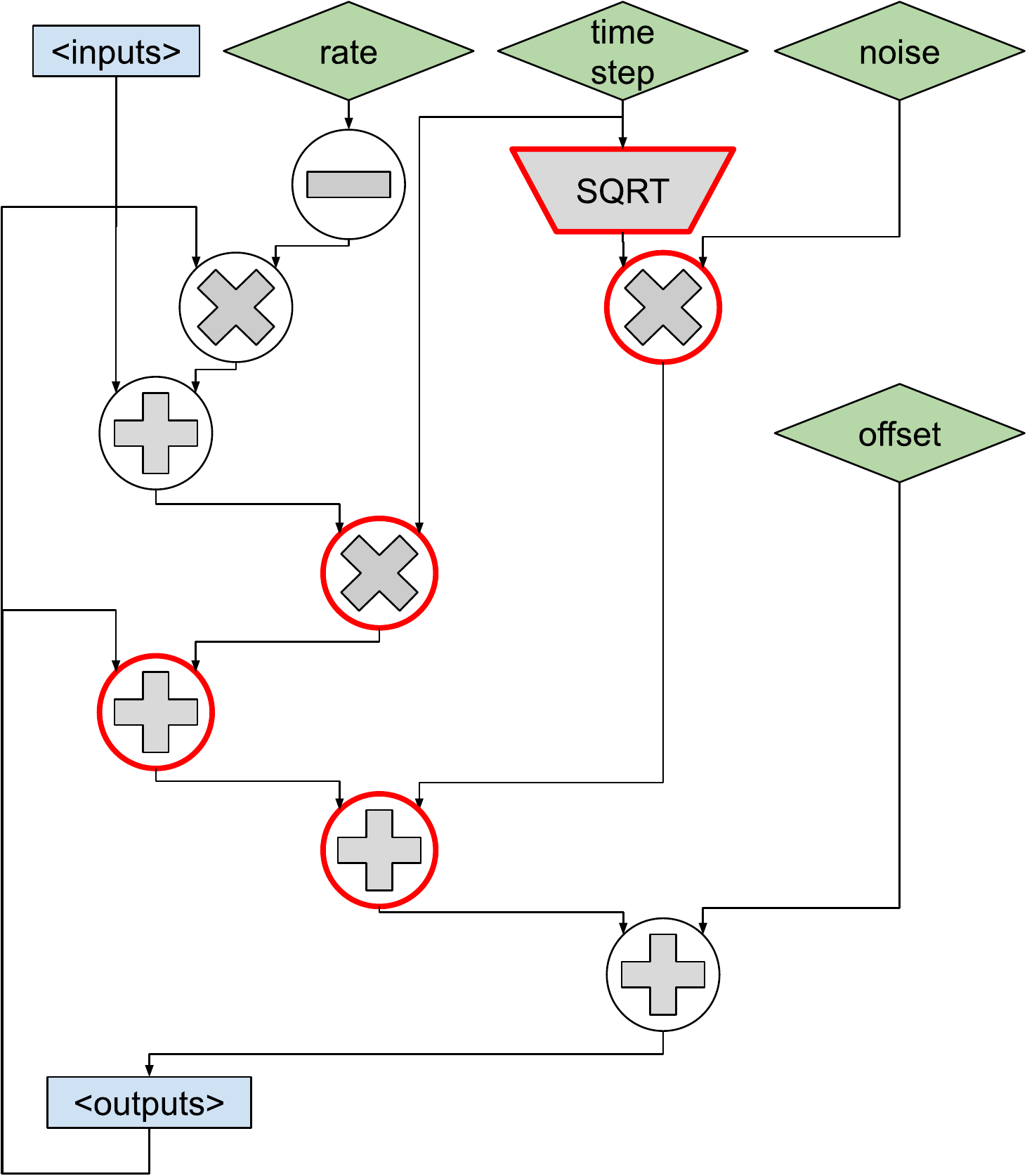}
}
\subfloat[Drift Diffusion Model (DDM)]{
\includegraphics[width=0.45\linewidth]{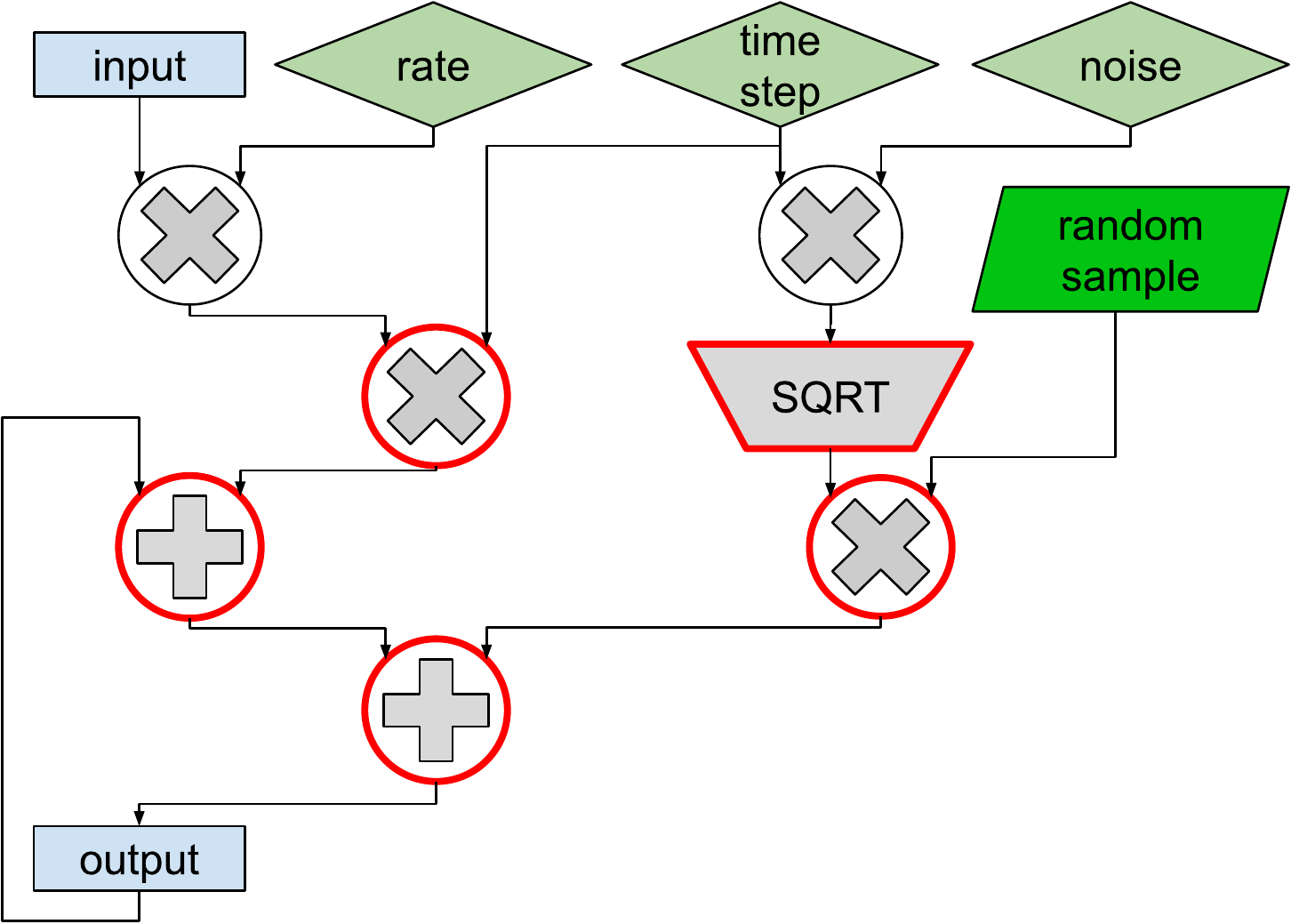}
}
\caption {Identical computation highlighted in red.
Setting $rate_{LCI} = 0$, $offset_{LCI} = 0$, $noise_{LCI} = N(0,1)$, $rate_{DDI} = 1$, and $noise_{DDI} = 1$, configures both functions to perform identical computation.}
\label{fig:clones-ddm-lca}
\end{figure}

To perform such analyses, we use LLVM's existing `FunctionComparator' framework to detect exactly equivalent functions. As an example of our analysis, consider two functions that use a Drift Diffusion Model (DDM) and a Leaky Competing Integrator (LCA)~\cite{bogacz2007extending} to simulate decision-making, respectively. DDM is used for two-choice decision making and has an analytical solution, while LCA is a multi-choice model. Figure~\ref{fig:clones-ddm-lca} shows that the underlying accumulation is very similar with identical sequence at the core of the computation, and the analysis is able to identify this correctly. Thus, it can notify the user when the LCA can be replaced with an analytical solution, saving thousands of model executions.

Aggressive inlining allows our methodology to work beyond functions and for entire models. Our analysis could find that a model for bistable perception of a Necker cube~\cite{necker1832lxi}, and its hand-tuned vectorized version are equivalent, even though they differ in their structure, number of nodes and the computation of each node. This is possible because our clone detection analysis works on the IR level,
independent of the original model's structure.

\begin{figure*}[ht]
\includegraphics[width=\linewidth]{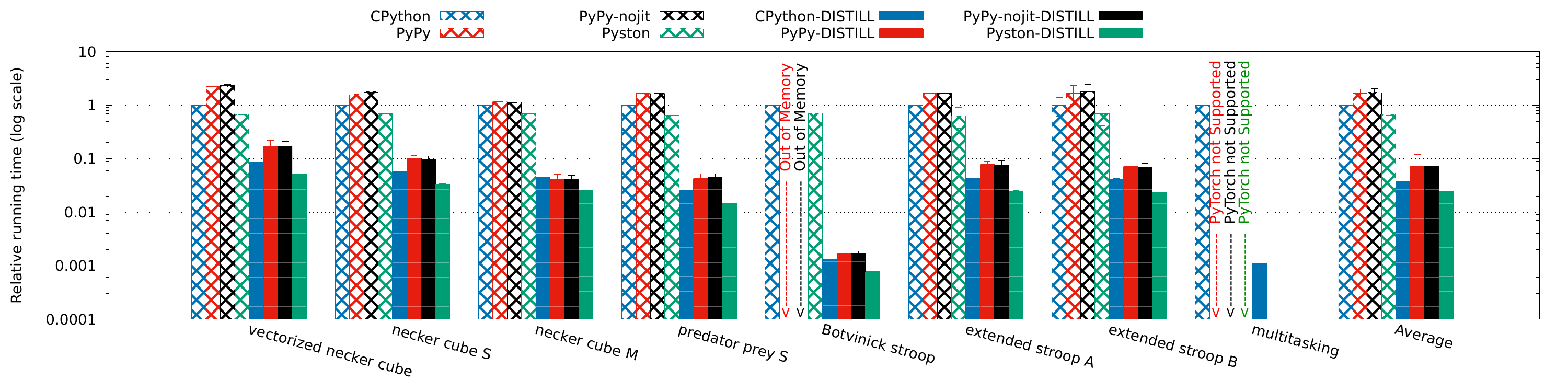}
\caption {Running times normalized to Python baseline.
{\em Botvinick stroop} did not complete when run using {\em PyPy3}.
{\em multitasking} uses {\em pytorch} which is not compatible with {\em PyPy3} or {\em Pyston}.}
\label{fig:performance}
\end{figure*}



\section{Experimental Setup}
\label{setup}

\newcommand{\pythonver}{Python-3.6.9}
\newcommand{\pypyver}{pypy3-7.3.2}
\newcommand{\pystonver}{pyston-2.0.0}
\newcommand{\cudaver}{CUDA 11.1}
\newcommand{\cpu}{Intel(R) Core(TM) i7-8700 CPU @ 3.20GHz}
\newcommand{\gpu}{GeForce GTX 1060 3GB VRAM}

We evaluate \sysname by using it to accelerate a selection of cognitive models designed in PsyNeuLink. We choose PsyNeuLink because it can integrate models from several frameworks. 
We use PsyNeuLink's python implementation as baseline and compare execution using \pythonver{} (\textit{CPython}), as well as two JIT enhanced implementations, \pystonver{} (\textit{Pyston}), and \pypyver{} (\textit{PyPy}). We also run PyPy without JIT compilation (\textit{PyPy-nojit}).
Since \sysname{} works with all python implementations we report \sysname{} model execution in all four environments.

\noindent
\textbf{Models for Testing}\\
\noindent
\textbf{Necker cube:} This model is used to simulate the perception of a subject when shown with bi-stable stimulus, typically the line drawing of a cube which can appear to either project out or into the screen~\cite{necker1832lxi}. The model contains one node per vertex and evaluates when the subject's perception oscillates between the two orientations due to gradual changes in the nodes' values. We evaluate three variants of the model: \textit{Necker cube S}, which is the model for a 3-vertex drawing, \textit{Necker cube M}, which uses a cube (8 vertices) and \textit{Vectorized Necker cube} which is a manually vectorized version of Necker cube M.


\noindent
\textbf{Predator-Prey:} We use 4 variants of the Predator-Prey model: S, M, L and XL that have 2, 4, 6 and 100 levels of attention per entity (prey, predator and player). Predator-Prey XL is representative of models that will be commonplace in future.

\noindent
\textbf{Botvinick Stroop Model:} The model simulates the conflict in the brain when processing the name of a color, and the ink color with which it is written~\cite{botvinick2001conflict}. The model calculates decision energy which is gradually changed over time by the stimulus, which is the colored word. The model is used to predict decision energy over time.

We also consider two extended versions of this model, \textit{Extended Stroop A} and \textit{Extended Stroop B} that add an additional task of finger pointing on top of the usual color naming. The model adds two DDM nodes (one for color naming, the other one for finger pointing) to the output of the Stroop model to produce a final decision. The versions A and B differ in how the inputs to the DDMs are computed and how its outputs determine the overall reward. Conceptually, they are different but they are computationally equivalent, as detected by \sysname{}.

\noindent
\textbf{Multitasking:} This model simulates conflict in representation when processing a combination of stimuli and goals. This model uses a neural network designed in PyTorch to give the color and shape of the stimulus. This information is passed to an LCA module designed in PsyNeuLink, which accumulates evidence to reach a decision. The model is run to obtain a distribution of response times and the histogram of correct and incorrect responses. This is a heterogeneous model that spans multiple execution environments: PyTorch and PsyNeulink.



The execution times are collected using {\em pytest-benchmark} package and we report the average running time and standard deviation error bars.
{\em pytest-benchmark} was configured to include two warmup runs before collecting the runtime data and therefore don't include the time to compile models, unless stated otherwise.

\noindent
\textbf{Infrastructure}\\
All experiments are done on an \cpu{} with 6 cores and 12 threads machine with 16GB of DDR4@2666Mhz RAM.
GPU experiments are run on a \gpu{} on the same machine using \cudaver{}.

\section{Performance Evaluation}
\label{eval}

Figure~\ref{fig:performance} shows the running time of the models with the different implementations viz., \textit{CPython}, \textit{PyPy}, \textit{PyPy-nojit} and \textit{Pyston} both with and without \sysname{}.
For the predator prey model, we only use the smallest variant in this chart, and analyze the remaining variants separately.
The execution times are normalized to those obtained from the \textit{CPython} implementation, and are plotted on a logarithmic scale. 

The execution times with \textit{PyPy} and \textit{PyPy-nojit} are 67\% and 71\% \textit{higher} than the standard \textit{CPython} execution times, on average.
\textit{Pyston} has a 43\% \textit{lower} execution time than \textit{CPython}, on overage.
The poor performance of \textit{PyPy} is surprising in that it claims to improve performance and lower memory usage. However, with, \textit{PyPy},  the \textit{Botvinick Stroop} model and the XL variant of the \textit{Predator Prey} model fail to complete after exhausting all 16$\,$GB of memory available on our test system. 

Pyston and PyPy cannot run the \textit{Multitasking} model because they do not support PyTorch.

When \sysname{} is used in these implementations, the execution times are 96\%, 93\%, 93\% and 98\% \textit{faster} than the standard execution for the respective environments, on average.
This translates to speedups of up to $26\times$, on average, and up to $\approx778\times$ for the Botvinick Stroop model.
The main reason is that the JIT compilers are designed for generic Python usage while \sysname{} exploits domain specific information for aggressive optimization.


\subsection{Scaling Model Sizes}

\begin{figure}[h]
\subfloat[Predator-Prey scaling]{
\includegraphics[width=.33\linewidth,valign=t]{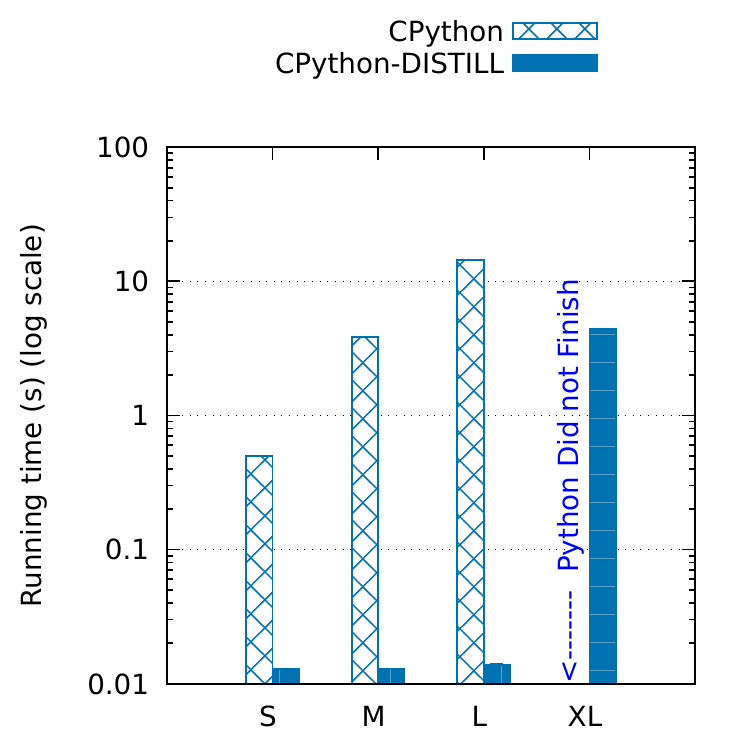}
\vphantom{\includegraphics[width=0.3\linewidth,valign=t]{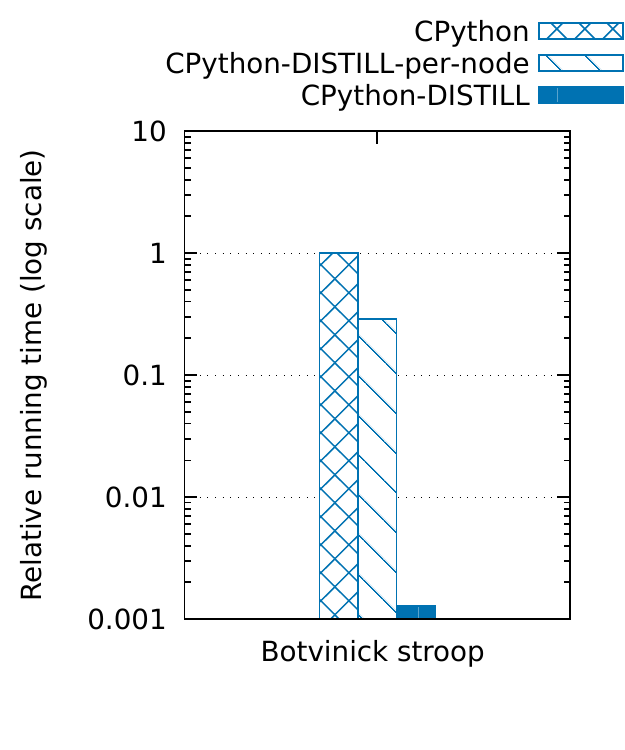}}
\label{fig:perf-scale}
}
\subfloat[Botvinick stroop model per-node][Botvinick stroop \\ model per-node]{
\includegraphics[width=.3\linewidth,valign=t]{figures/perf-stroop-per-node.pdf}
\label{fig:perf-per-node}
}
\subfloat[Predator-Prey XL parallel]{
\includegraphics[width=.27\linewidth,valign=t]{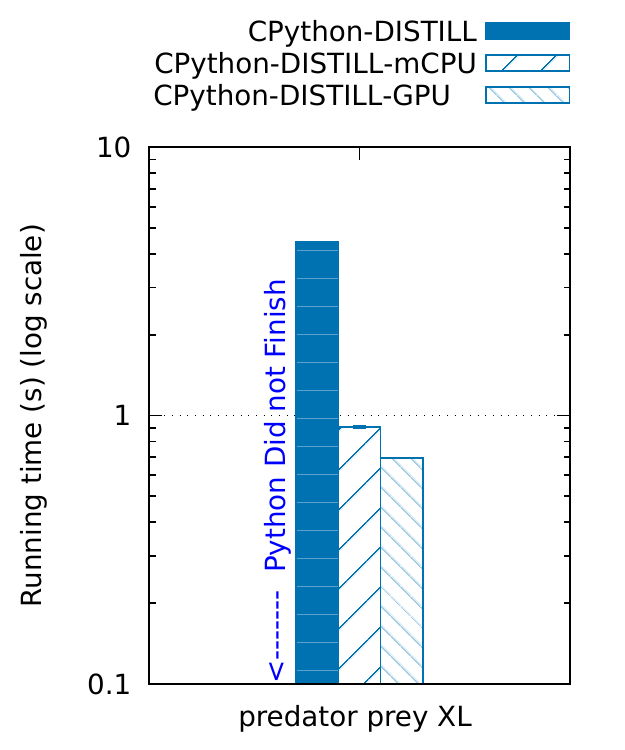}
\vphantom{\includegraphics[width=0.3\linewidth,valign=t]{figures/perf-stroop-per-node.pdf}}
\label{fig:perf-parallel}
}

\caption{a) shows how \sysname can accelerate models as they are scaled. b) shows the importance of accelerating the entire model. c) compares different parallel evaluations of Predator-Prey XL in the model, GPU vs. 6C/12T CPU. }
\end{figure}

We use the Predator-Prey model to present how \sysname{} can accelerate models as we increase their computational intensity.
Figure~\ref{fig:perf-scale} shows the execution time of the four variants of the predator-prey model for the \textit{CPython} environment and when using \sysname{}.
Recall that the variants have 2, 4, 6 and 100 attention levels per entity corresponding to 8, 64, 216 and 1,000,000 evaluations, respectively. 

From the figure,the runtime for the smaller models (S, M and L) with \sysname{} remains nearly the same while the baseline takes an order of magnitude longer time for each step up in the number of attention levels.
With XL, the model does not complete even after 24 hours using \textit{CPython} alone, while with \sysname{} the model finishes execution in about 4 seconds.
Despite the number of evaluations increasing by 4600$\times$ from L to XL, the running time with \sysname{} only goes up by $\approx$330$\times$ from $\approx$0.02$\,$s to $\approx$4.4$\,$s.
This shows that \sysname{} works very well even as we scale models. Importantly, with more realistic cognitve models, \sysname{} is the only choice to keep execution times reasonable.

\subsection{Importance of Model-Wide Optimizations}

We highlight the importance of model-wide optimizations that span the nodes and the scheduling logic of PsyNeuLink.
Figure~\ref{fig:perf-per-node} shows the normalized running time of the \textit{Botvinick Stroop} model when \sysname{} is used to generate optimized code per node but with optimizations not crossing node boundaries (the \textit{CPython-\sysname-per-node} design).
This also means that execution switches between the scheduling logic in Python and the compiled model nodes. Compared to \textit{CPython} execution, 
\sysname{}-per-node compilation and model-wide compilation (default \sysname{}) result in 70\% and 99.8\% performance improvements, translating to $3.4\times$ and $778\times$ speedups respectively.
This shows the tremendous impact of model-wide optimizations that \sysname{} enables.

\subsection{Parallel and GPU Execution}

Figure~\ref{fig:perf-parallel} shows the  execution times in seconds when the predator-prey mode XL (the largest of our tested models) is run on a 12-threaded multicore CPU and a GPU.
Recall that \sysname{} automatically generated parallel code for both systems.

As mentioned earlier, this model did not complete execution in the standard environment. The single thread, multithread CPU (mCPU) and GPU executions result in execution times of 4.4$\,$s, 0.9$\,$s and 0.7$\,$s corresponding to speedups of 4.9$\times$ and 6.3$\times$, respectively, over the serial execution. 

One reason for the less-than-ideal speedup on the GPUs is that our implementation replicates the state of the PRNGs used in each thread. Each thread uses 3 PRNGs to obtain the observed X and Y coordinates of the predator, prey and the player, resulting in a total PRNG state of nearly 7.5$\,$kB per thread. Such a large storage causes pressure on the GPU memory hierarchy, resulting in slower execution. 

While we could have used GPU-friendly PRNGs, we did not do so because the choice of a PRNG affects the values of the model output~\cite{click:quality}. There is no theoretical analysis on how  a new PRNG would impact the predator-prey model.

\begin{figure}[ht]
\includegraphics[width=\linewidth]{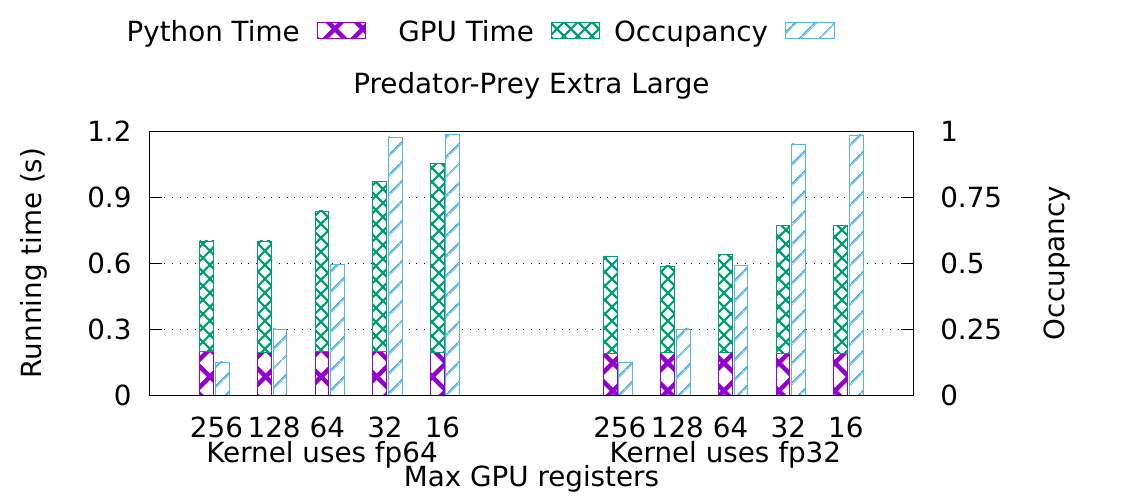}
\caption {GPU running time using different restrictions on available register space.
Total size of private (per-thread) data is 18.5kB for double precision variant, and 15.5kB for single precision.}
\label{fig:gpu-performance}
\end{figure}

To confirm that memory pressure is the cause for the suboptimal GPU performance, we run additional studies. Figure~\ref{fig:gpu-performance} shows the execution time and occupancy (defined as the ratio of number of active threads to the maximum) in fp32 and fp64 modes, when the maximum allowed registers per thread is throttled. As fewer registers are used, occupancy increases but execution time increases too. Additionally, fp32, which typically provides up to 32x the computational throughput of fp64, results in nearly the same speedup. This indicates that the bottleneck is not in compute, and is in fact in memory. 

\subsection{Cost of Compilation}

\begin{figure}[h]
\includegraphics[width=\linewidth]{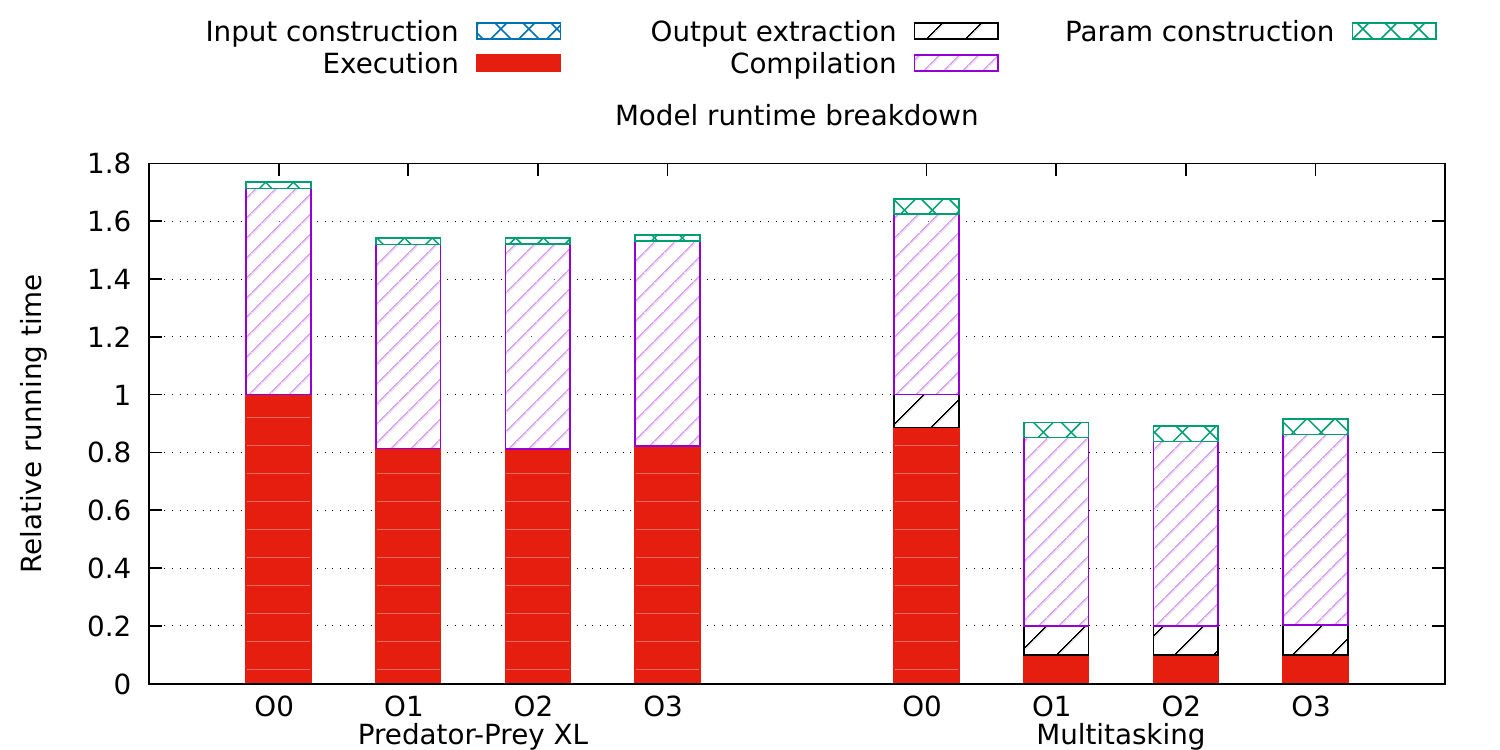}
\caption {Running time breakdown of models.}
\label{fig:breakdown}
\vspace{-5mm}
\end{figure}

Figure~\ref{fig:breakdown} shows the cost of dynamic compilation for two of our large models, the XL variant of predator-prey and the Multitasking model. The values are normalized to the execution time of the Predator-Prey model with O0 optimizations. The chart also shows the fraction of time to convert inputs and outputs to arrays, and parameters to statically defined structures. Even though compilation times are significant, they are amortized because the models are typically run hundreds to thousands of times after compilation.

\section{Related Work}

\sysname{} builds on, connects, and extends proven approaches across different fields.

On the high level, JIT based impelmentations of high level languages, like PyPy3~\cite{pypy} and Pyston~\cite{pyston} are the closest to \sysname{}.
However, unlike PyPy3 and Pyston, \sysname{} doesn't need to support the Python language in its entirety can and focuses only on parts required by the cognitive models.
We show that eschewing the full generality of Python can be exploited for significant performance benefit,
and enables usable analytical feedback to modeler.

Our focus on compiling only a subset of Python language resembles Numba~\cite{numba,numba-limitations}, a popular approach to achieve compiled performance while using Python.
Numba, however, cannot be directly applied to \pnlname{} codebase.
It lacks the domain specific information to understand high level modeling semantics and avoid unsupported language constructs.
Unlike \sysname{}, Numba pushes the burden of eliminating expensive dynamic constructs the the PsyNeuLink developers and cognitive modelers.

\sysname{} is not the first to notice the significant opportunity in applying domain specific knowledge to Python.
Spiegelberg et al.~\cite{tuplex} apply domain specific knowledge of data types and hot paths to construct efficient data processing pipelines.
The are able to extract and compile expected hot paths with exception checks that fall back to fully interpreted Python.
Unlike Tuplex, \sysname{} benefits from domain specific knowledge that makes full python fallback unnecessary.
This enables us to represent models in LLVM IR completely and use the same IR for analysis and not just performance.

Weld~\cite{weld} similarly exploits the benefits of representing complete programs in a single IR to achieve good performance.
Unlike Weld, \sysname{}, doesn't introduce a new IR, but relies on an industry standard IR.
We also do not require third parties to provide an IR representation and instead rely on \pnlname{}'s ability to import models from different modeling environments.

The model analysis presented in this paper exploits the link to software engineering techniques.
Although \sysname{} explored the utility of direct model comparison in their IR form, this representation opens the door to other clone detection techniques known from the software engineering domain~\cite{Gabel2008-ya,Keivanloo2012-ej,Schugerl2011-jk}.
We believe these approaches to be fully applicable to cognitive models,
and have potential to further benefit the modeling effort by affording richer comparison between models.
Some have even explored applicability to block modeling environments~\cite{Alalfi2012-ke:models-are-code}.

\section{Conclusions}

Cognitive models are vital in understanding and replicating the processes behind human cognition. {\sysname} examines the role of compilers in supporting robust and high-performance modeling of these types of cognitive processes.
Beyond offering large performance improvements necessary to support complex and higher fidelity cognitive models,
we present the suitability of compiler analyses for cognitive modeling analyses.
We propose and implement modifications to a production compiler suite to provide rich feedback to cognitive modelers.
All our contributions are part of open-source projects and released for public use.

\bibliographystyle{ACM-Reference-Format}
\bibliography{references}

\end{document}